\documentclass[aps,prl,twocolumn,floatfix,superscriptaddress,longbibliography]{revtex4-2}

\usepackage[latin9]{inputenc}
\usepackage{amsmath}
\usepackage{amssymb}
\usepackage{graphicx}
\usepackage{esint}
\usepackage{epstopdf}
\usepackage{braket}
\usepackage{color}
\usepackage[colorlinks=true,citecolor=blue,linkcolor=blue,urlcolor=blue]{hyperref}
\usepackage{breakurl}
\usepackage{tikz-feynman}
\usepackage{bibunits}
\usepackage{mathtools}
\usepackage{stackengine}
\usepackage{dsfont}
\usepackage{soul}
\usepackage{multirow}
\usepackage{placeins}

\begin{document}

\title{Entanglement and Bell nonlocality with bottom-quark pairs at hadron colliders}

\author{Yoav Afik}
\email{yoavafik@gmail.com}
\affiliation{Enrico Fermi Institute, University of Chicago, Chicago, Illinois 60637, USA}

\author{Yevgeny Kats}
\email{katsye@bgu.ac.il}
\affiliation{Department of Physics, Ben-Gurion University, Beer-Sheva 8410501, Israel}

\author{Juan Ram\'on Mu\~noz de Nova}
\email{jrmnova@fis.ucm.es}
\affiliation{Departamento de F\'isica de Materiales, Universidad Complutense de Madrid, E-28040 Madrid, Spain}

\author{Abner Soffer}
\email{asoffer@tau.ac.il}
\affiliation{School of Physics and Astronomy, Tel Aviv University, Tel Aviv 69978, Israel}

\author{David Uzan}
\email{daviduz@post.bgu.ac.il}
\affiliation{Department of Physics, Ben-Gurion University, Beer-Sheva 8410501, Israel}

\begin{abstract}
It has been shown that entanglement and Bell nonlocality, which are key concepts in Quantum Mechanics, can be probed in high-energy colliders via processes of fundamental particle scattering. 
In fact, the ATLAS and CMS collaborations have measured entanglement using top-quark pairs produced in proton-proton collisions at the LHC. 
Recently, it was shown that spin correlations can be measured in pairs of bottom quarks at the LHC, despite the fact that bottom quarks, unlike top quarks, hadronize before decaying. 
Here, we demonstrate that quantum correlations can also be studied using bottom-quark pairs, and analyze the feasibility of the observation of entanglement and Bell nonlocality in several collider experiments.
Given the low mass of the bottom quark relative to typical energies accessible at the LHC, many of the bottom-quark pairs are in the ultrarelativistic regime, where they can exhibit strong spin entanglement. We find that entanglement of bottom-quark pairs may be measurable even with the LHC Run 2 data, especially with the CMS $B$ parking dataset, while observation of Bell nonlocality may become feasible at the high-luminosity phase of the LHC.
\end{abstract}

\maketitle

\textit{Introduction.---}Entanglement is a fundamental property of Quantum Mechanics (QM)~\cite{Einstein:1935rr,Schrodinger1935}: if a pair of particles is entangled, the state of the system cannot be described by specifying the state of each particle separately.
A most remarkable manifestation of entanglement is the violation of Bell-type inequalities~\cite{Bell:1964kc}, which addresses the counterintuitive absence of local realism.
Both concepts are at the heart of QM, and have been probed in a variety of systems at vastly different scales~\cite{Aspect1982,Hagley1997,Steffen2006,Pfaff2013,Belle:2007ocp,Julsgaard2001,Lee2011,Ockeloen2018,Storz:2023jjx}.

Particle colliders, such as the LHC, offer a unique setup for these experiments, since they allow studying QM at the highest energies accessible to us. 
The feasibility of Bell tests at colliders was originally discussed in Refs.~\cite{Tornqvist:1980af,Tornqvist:1986pe,Privitera:1991nz,Abel:1992kz}.
Recently, it has been shown that top-quark pairs provide a paradigmatic high-energy platform to study quantum correlations~\cite{Afik:2020onf,Fabbrichesi:2021npl,Severi:2021cnj,Afik:2022kwm,Aguilar-Saavedra:2022uye,Afik:2022dgh,Ashby-Pickering:2022umy,Aguilar-Saavedra:2023hss,Cheng2023,Dong:2023xiw,Aguilar-Saavedra:2024hwd,Aguilar-Saavedra:2024fig,White:2024nuc}. 
This is because the large top-quark mass ($m_t \approx 173$~GeV) results in an extremely short lifetime ($\tau_t \approx 5 \times 10^{-25}$\,s), significantly shorter than the typical time for hadronization ($1/\Lambda_{\rm QCD} \approx 3 \times 10^{-24}$\,s) or spin decorrelation ($m_t/\Lambda_{\rm QCD}^2 \approx 2 \times 10^{-21}$\,s).
Therefore, the information on the spins of the top quarks is propagated directly to the decay products, from where the polarizations and spin correlations of the top-quark pair are reconstructed.
Indeed, recent analyses by the ATLAS and CMS collaborations have observed  entanglement between top quarks~\cite{ATLAS:2023fsd,CMS:2024pts,CMS:2024zkc}. Neutrinos~\cite{Blasone:2007vw,Formaggio:2016cuh,Ming2020,Blasone:2021cau}, $\tau$~leptons~\cite{Fabbrichesi2022,Altakach:2022ywa,Ehataht:2023zzt,Fabbrichesi:2024wcd}, or massive gauge bosons~\cite{Barr2022,Barr2022B,Ashby-Pickering:2022umy,Aguilar2023,Aguilar2023a,Bernal:2023ruk,Morales:2023gow,Fabbri:2023ncz} offer parallel avenues, as reviewed in Ref.~\cite{Barr:2024djo}. 
Quantum information techniques can be used to probe nonperturbative features of quantum chromodynamics (QCD)~\cite{Baker:2017wtt,Tu2020,Gong:2021bcp,Florio:2023dke,Barata:2023jgd,Hentschinski:2024gaa,Von2025}.

Even though lighter quarks hadronize, it has been recently shown that their polarizations and spin correlations can also be measured using their hadronization products, albeit with some loss of precision~\cite{Kats:2023zxb}. In particular, bottom-antibottom-quark ($b\bar{b})$ pairs were found to be promising.
Since the bottom-quark mass is only $m_b\approx 5$~GeV, $b\bar b$ pairs are copiously produced at the LHC in the ultrarelativistic regime, $M_{b \bar b} \gg m_b$.
This makes the $b\bar b$ system particularly attractive for the study of this regime.

In this letter we show that the LHC experiments ATLAS~\cite{ATLAS:2008xda}, CMS~\cite{CMS:2008xjf}, and LHCb~\cite{LHCb:2008vvz}, with either standard or special trigger paths, are promising for detection of entanglement and Bell nonlocality in $b\bar b$ pairs. Some measurements can be pursued with data already collected, while others will become feasible at the high-luminosity phase of LHC (HL-LHC).

\textit{General formalism.---}A $b \bar b$ pair forms a bipartite system of two spin-$1/2$ particles. As such, it is described by the density matrix 
\begin{equation}\label{eq:GeneralBipartiteStateRotations}
\rho=\frac{I_4+\sum_{i}\left(B^{+}_i\sigma^i\otimes I_2+B^{-}_i I_2\otimes\sigma^i \right)+\sum_{i,j}C_{ij}\sigma^{i}\otimes\sigma^{j}}{4} ,
\end{equation}
where $I_n$ is the $n\times n$ identity matrix, $B^{\pm}_i$ are the components of the Bloch vectors $\mathbf{B}^{\pm}$ that represent the bottom/antibottom-quark polarizations, $C_{ij}$ are the elements of the spin-correlation matrix $\mathbf{C}$, and $\sigma^i$ are the Pauli matrices.

At hadron colliders, at leading order (LO) in QCD, bottom-quark pairs are produced from quark-antiquark ($q \bar q \to b \bar b$) or gluon-gluon ($gg \to b \bar b$) interactions. 
Kinematically, a $b \bar b$ pair in its center-of-mass (COM) frame is specified by the invariant mass $M_{b \bar b}$ and the $b$-quark direction $\hat{k}$.
For fixed $(M_{b \bar b},\hat{k})$, the spin quantum state of the $b \bar b$ pair is described by the density matrix $\rho(M_{b\bar{b}},\hat{k})$, which is the weighted sum of the density matrices $\rho^{I}(M_{b\bar{b}},\hat{k})$ arising from the initial states $I=q \bar q,gg$~\cite{Afik:2020onf,Afik:2022kwm},
\begin{equation}\label{eq:partonicstates}
\rho(M_{b\bar{b}},\hat{k})=\sum_{I=q\bar{q},gg} w_I(M_{b\bar{b}})\rho^{I}(M_{b\bar{b}},\hat{k}) \,,
\end{equation}
where the weights $w_I$ are determined by the parton distribution functions (PDFs) that effectively describe the content and structure of the colliding hadrons. 
Gluon radiation from the final-state $b$ quarks is not expected to affect their spin significantly~\cite{Neubert:1996wg,Korner:1993dy}.
More generally, higher-order QCD corrections to $b\bar b$ spin correlations are likely to be small, as was found for $t\bar t$ spin correlations~\cite{Bernreuther:2004jv}.

The orthonormal basis customarily used to evaluate $B^{\pm}_i$ and $C_{ij}$ is the helicity basis (e.g., Ref.~\cite{Baumgart:2012ay}), defined in the $b\bar b$ COM frame in terms of the vectors $\{\hat{k},\hat{n},\hat{r}\}$. 
Here $\hat{r}=(\hat{p}-\cos\Theta\,\hat{k})/\sin\Theta$ and $\hat{n}=\hat{r}\times\hat{k}$, where $\hat{p}$ is the proton-beam axis, and $\cos\Theta=\hat{k}\cdot\hat{p}$. At LO in QCD, the $b \bar b$ quantum state is unpolarized ($B^{\pm}_i=0$), and the only nonvanishing spin-correlation matrix elements are $C_{kk}$, $C_{nn}$, $C_{rr}$, $C_{rk}=C_{kr}$. The analytical expression for $\rho^I$ is the same as in the $t\bar t$ case~\cite{Bernreuther1994,Bernreuther1998,Baumgart:2012ay}, and solely function of $(\beta,\cos\Theta)$, with $\beta=\sqrt{1-4m^2_b/M^2_{b\bar{b}}}$ the bottom-quark velocity in the COM frame.

The polarization and spin correlations for $b\bar b$ pairs can be measured using events in which the $b$ quarks hadronize into baryons. The lightest, most commonly produced $b$-baryon is the $\Lambda_b$, which in the simple quark model contains a spin-singlet, isospin-singlet combination of the light quarks $u$ and $d$ in addition to the $b$ quark, which carries the baryon spin. Since $m_b \gg \Lambda_{\rm QCD}$, $\Lambda_b$ baryons are expected to carry a large fraction of the original $b$-quark polarization~\cite{Mannel:1991bs,Ball:1992fw,Falk:1993rf,Galanti:2015pqa}. The baryon polarization can be measured from the kinematic distribution of its decay products. This has been done, although with large statistical uncertainties, in $Z\to b\bar b$ decays at LEP, using semileptonic decays of the $\Lambda_b$~\cite{ALEPH:1995aqx,OPAL:1998wmk,DELPHI:1999hkl}.

Spin-correlation measurements can be performed with both the $\Lambda_b$ and $\overline\Lambda_b$ decaying semileptonically via $\Lambda_b \to X_c\ell^-\bar\nu_\ell$, where $X_c$ denotes a charmed state containing a baryon, usually the $\Lambda_c^+$. 
The angular distribution of the neutrinos from the two semileptonic decays is approximately~\cite{Kats:2023zxb}
\begin{equation}
\frac{1}{\sigma}\frac{d\sigma}{dx_{ij}} =
\frac{1}{2}\left(1 - c_{ij}x_{ij}\right) \ln\left(\frac{1}{|x_{ij}|}\right) ,
\label{eq:ang-distrib}
\end{equation}
where $x_{ij} = \cos\theta^+_i\cos\theta^-_j$, the angles $\theta^+_i$ ($\theta^-_j$) describe the directions of the antineutrino (neutrino) from the $\Lambda_b$ ($\overline\Lambda_b$) decay in the respective parent rest frames with respect to the $i$ ($j$) axis, and
\begin{equation}
c_{ij} = \alpha^2 r_i r_j C_{ij} \,.
\label{eq:HadToParton Relation}
\end{equation}
Here, $\alpha \simeq 1$ is the spin analyzing power of the (anti)neutrino in the semileptonic $\Lambda_b$ decay. The factors $r_i$ and $r_j$ are the longitudinal ($r_L$, for the $\hat k$ axis) or transverse ($r_T$, for the $\hat n$ and $\hat r$ axes) polarization retention factors~\cite{Falk:1993rf,Galanti:2015pqa}. They describe the fraction of the original quark's longitudinal or transverse polarization that is retained in the baryon and are rough approximations of the spin-dependent fragmentation functions (e.g., Refs.~\cite{Metz:2016swz,Chen:1994ar}). Their values are expected to be roughly in the ranges $0.4 \lesssim r_L \lesssim 0.8$, $0.5 \lesssim r_T \lesssim 0.8$~\cite{Galanti:2015pqa} (see Supplemental Material for additional details~\cite{SM}), where the dominant polarization loss arises from $\Lambda_b$ baryon production in $\Sigma_b^{(\ast)}$ decays~\cite{Falk:1993rf,Galanti:2015pqa}. An approximate combination of the LEP measurements~\cite{ALEPH:1995aqx,OPAL:1998wmk,DELPHI:1999hkl} gives $r_L = 0.47 \pm 0.14$. It is also possible to measure $r_L$ with competitive level of precision with the data already available in ATLAS and CMS, using the highly polarized $b$ quarks from $pp \to t\bar t$ events~\cite{Galanti:2015pqa}. 
In addition, both $r_L$ and $r_T$ can be measured in $pp \to b\bar b$ events~\cite{Kats:2023zxb}. This can be done independently of the entanglement measurement using a dedicated control region of phase space where significant entanglement is not expected while some of the elements $C_{ij}$ are sizable.

Measuring $C_{ij}$ requires reconstructing the momenta of the two $b$-quarks, the $\Lambda_b$ baryons, and the neutrinos. This can be approximately done as outlined in Refs.~\cite{Galanti:2015pqa,Dambach:2006ha,LHCb:2015eia,LHCb:2020ist,Ciezarek:2016lqu,Dambach:2009wda}. The approximations involved will need to be accounted for in interpreting the data and will lead to a reduction in sensitivity, the evaluation of which is beyond the scope of the current work.

\textit{Entanglement and Bell nonlocality.---}A quantum state in a composite Hilbert space is said to be separable if it can be written as a convex sum of product states. For a bipartite system such as a $b\bar b$ pair, separability implies that
\begin{equation}\label{eq:Separability}
\rho=\sum_n p_n \rho^{b}_{n}\otimes\rho^{\bar{b}}_{n}\,,~\sum_n p_n=1\,,~p_n\geq 0\,,
\end{equation}
where $\rho^{b}_{n}$, $\rho^{\bar{b}}_{n}$ are density matrices in the bottom, antibottom-quark subspaces. Entanglement is defined as the nonseparability of a quantum state, i.e., that the state cannot be written in this form. A quantitative measurement of entanglement is provided by the concurrence $\mathcal{C}$~\cite{Wooter1998}, which satisfies $0\leq \mathcal{C}\leq 1$, where $\mathcal{C}>0$ is a sufficient and necessary condition for entanglement and  $\mathcal{C}=1$ corresponds to a maximally entangled state. At LO QCD, the concurrence is given by $\mathcal{C}=\max(\Delta,0)$~\cite{Afik:2022kwm}, where
\begin{equation}\label{eq:EntanglementWitness}
    \Delta\equiv \frac{-C_{nn}+|C_{kk}+C_{rr}|-1}{2} \,.
\end{equation}
In general, $\Delta>0$ is a sufficient condition for entanglement and can then be used as an entanglement marker.

Highly entangled states may violate Bell inequalities. These states are denoted as Bell nonlocal and, throughout this work, we generally refer to this property as Bell nonlocality. Although current collider setups do not allow for proper Bell tests using particle spins, it is still possible to observe quantum states that exhibit Bell nonlocality. For spin-1/2 particles, a useful form of Bell inequality is the Clauser-Horne-Shimony-Holt (CHSH) inequality~\cite{Clauser1969}. A quantum state can violate the CHSH inequality \textit{iff} $\mu_1+\mu_2>1$, with $0\leq \mu_2 \leq \mu_1 \leq 1$ the two largest eigenvalues of $\mathbf{C}^{\textrm{T}}\mathbf{C}$~\cite{Horodecki1995}. In practice, a simple sufficient condition for CHSH violation is $\mathcal{V}>0$, where~\footnote{We have checked that other procedures to evaluate the Bell nonlocality marker, such as directly averaging $\mathcal{V}$ in phase space or alternative basis choices, do not substantially improve the significance of the observation.}
\begin{equation}\label{eq:CHSHviolation}
    \mathcal{V}\equiv C^2_{kk}+C^2_{rr}-1\leq \mu_1+\mu_2-1 \,.
\end{equation}
This observable is expected to accurately capture the Bell nonlocality of the $b \bar b$ quantum state in the ultrarelativistic regime, $M_{b\bar b}\gg 2m_b$, in which $C_{kr}\simeq 0$ (so then $\mathbf{C}$ is diagonal), and $C^2_{kk},C^2_{rr}>C^2_{nn}$ \cite{Baumgart:2012ay}.

\begin{figure}[tb!]
    \includegraphics[width=0.49\textwidth]{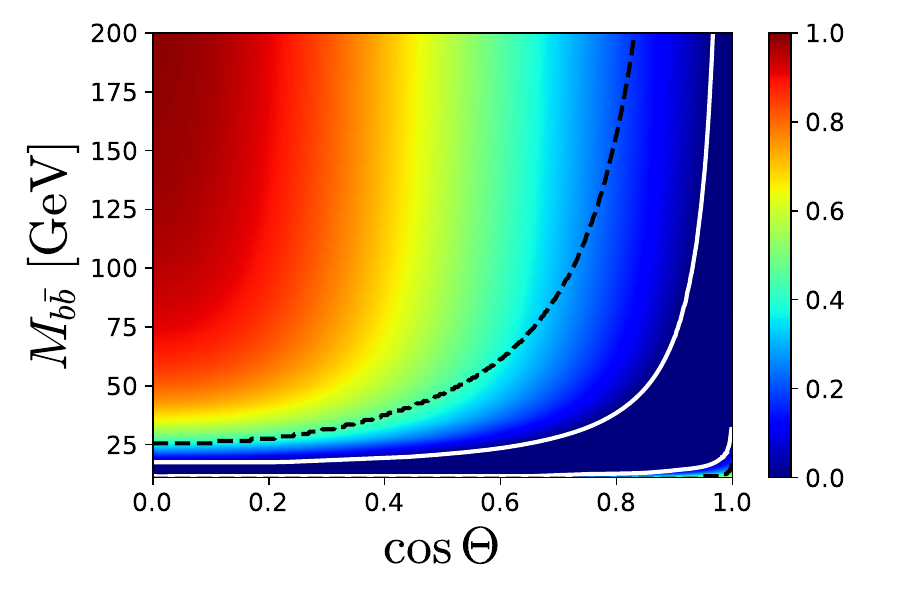} 
\caption{Concurrence $\mathcal{C}$ of the $b\bar b$ quantum state $\rho(M_{b\bar{b}},\hat{k})$ for $pp$ collisions at a COM energy of $\sqrt{s}=13$~TeV, as a function of the invariant mass $M_{b\bar{b}}$ and the production angle $\Theta$ of the bottom quark in the $b\bar b$ COM frame, computed at LO.
Entanglement (Bell nonlocality) takes place in the regions outside the solid white (dashed black) lines.} 
\label{fig:Concurrence}
\end{figure}

Figure \ref{fig:Concurrence} shows the concurrence of the $b \bar b$ spin quantum state at the LHC, computed analytically from Eq.~(\ref{eq:partonicstates}) with the NNPDF3.0 LO PDFs~\cite{NNPDF:2014otw}, as a function of $M_{b\bar{b}}$ and $\cos\Theta$; all relevant quantities are even functions of $\cos\Theta$. Entanglement (Bell nonlocality) takes place in the regions outside the solid white (dashed black) lines. Due to the dominance of the $gg$ production channel, we can understand the entanglement structure in similar terms to that of $t \bar t$ production~\cite{Afik:2020onf,Afik:2022kwm}. Close to threshold, $M_{b\bar b}\simeq 2m_b$, the $b \bar b$ system is in a spin singlet, maximally entangled. However, in contrast to the $t\bar t$ case, the threshold region is small relative to the range of $M_{b\bar{b}}$ achievable at the LHC. In the ultrarelativistic regime, the $b \bar b$ are in a maximally entangled spin-triplet state for transverse production ($\cos\Theta\simeq 0$).

\textit{Feasibility study.---}We study the feasibility of measuring $\Delta$ and $\mathcal{V}$ at the LHC with Run~2 data, as well as at the future HL-LHC~\cite{ZurbanoFernandez:2020cco,CERN-LHCC-2021-012}, assuming that similar triggers will be employed there. We consider the datasets of ATLAS/CMS (using the ATLAS parameters for our estimates) and LHCb, as well as the Run~2 CMS $B$ parking dataset~\cite{Bainbridge:2020pgi,CMS:2024zhe,CMS:2024syx,CMS:2024ita}. We use the estimated signal and background event counts from Ref.~\cite{Kats:2023zxb} for ATLAS, and perform additional simulations  to estimate the event counts for LHCb and the CMS $B$ parking dataset.
We also compute the spin correlations for all cases.

\begin{table*}[ht!]
\centering
\begin{tabular}{c|c c c c c c c c c c c c c c c}\hline\hline\noalign{\smallskip}
    & $\sigma\epsilon_{\mu\mu}$~[pb] & $\mathcal{L}~[\text{fb}^{-1}]$ & $N$ & $C_{kk}$ & $C_{rr}$ & $C_{nn}$ & $\Delta$ & $\mathcal{V}$ & $r_L$ & $\sigma_{\Delta}^\text{stat}$ & $\sigma_{\mathcal{V}}^\text{stat}$ & $\dfrac{\Delta}{\sigma_{\Delta}^\text{stat}}$ & $\dfrac{\mathcal{V}}{\sigma_{\mathcal{V}}^\text{stat}}$&$\dfrac{\Delta}{\sigma_{\Delta}^\text{tot}}$ &$\dfrac{\mathcal{V}}{\sigma_{\mathcal{V}}^\text{tot}}$\\\noalign{\smallskip}\hline
    & \multicolumn{15}{c}{\textbf{Run~2,} $\sqrt{s}=13$~TeV} \\\hline
    \multirow{2}{*}{ATLAS} & \multirow{2}{*}{$1.9 \times 10^4$} & \multirow{2}{*}{$140$} & \multirow{2}{*}{$2.7 \times 10^4$}&\multirow{2}{*}{$0.94$}&\multirow{2}{*}{$0.57$}&\multirow{2}{*}{$-0.56$}&\multirow{2}{*}{$0.54$} & \multirow{2}{*}{$0.21$}&0.75&0.14&0.33&$3.9$&$0.6$&$3.1$&$0.6$\\
&&&&&&&&&0.45&0.23&0.78&$2.3$&$0.3$&$2.1$&$0.3$\\
\hline
\multirow{2}{*}{LHCb, $\Delta > 0.2 $} & \multirow{2}{*}{$3.9 \times 10^6$} & \multirow{2}{*}{$5.7$} & \multirow{2}{*}{$1.8 \times 10^4$}&\multirow{2}{*}{$0.55$}&\multirow{2}{*}{$0.67$}&\multirow{2}{*}{$-0.56$}&\multirow{2}{*}{$0.39$} & \multirow{2}{*}{$-0.24$}&0.75&0.17&0.34&$2.2$&$-0.7$&$2.0$&$-0.7$\\
&&&&&&&&&0.45&0.29&0.62&$1.3$&$-0.4$&$1.3$&$-0.4$\\
\hline
\multirow{2}{*}{CMS $B$ parking, $\Delta > 0.2$ \;} & \multirow{2}{*}{$7.9 \times 10^5$} & \multirow{2}{*}{$41.6$} & \multirow{2}{*}{$1.8 \times 10^5$}&\multirow{2}{*}{$0.76$}&\multirow{2}{*}{$0.63$}&\multirow{2}{*}{$-0.59$}&\multirow{2}{*}{$0.49$} & \multirow{2}{*}{$-0.03$}&0.75&0.055&0.120&$8.9$&$-0.3$&$4.4$&$-0.3$\\
&&&&&&&&&0.45&0.092&0.256&$5.3$&$-0.1$&$3.6$&$-0.1$\\
\hline
    & \multicolumn{15}{c}{\textbf{HL-LHC,} $\sqrt{s}=14$~TeV}\\\hline
\multirow{2}{*}{ATLAS, $\mathcal{V}>0.3$} & \multirow{2}{*}{$9.9 \times 10^4$} & \multirow{2}{*}{$3000$} & \multirow{2}{*}{$1.0 \times 10^6$}&\multirow{2}{*}{$0.91$}&\multirow{2}{*}{$0.85$}&\multirow{2}{*}{$-0.83$}&\multirow{2}{*}{$0.79$} & \multirow{2}{*}{$0.55$}&0.75&0.02&0.06& $> 10$&$8.7$&$4.9$&$4.3$\\
&&&&&&&&&0.45&0.04&0.13& $> 10$&$4.3$&$4.9$&$3.3$\\
\hline
\multirow{2}{*}{LHCb, $\mathcal{V} > 0.3 $} & \multirow{2}{*}{$4.3 \times 10^6$} & \multirow{2}{*}{$300$} & \multirow{2}{*}{$8.2 \times 10^4$}&\multirow{2}{*}{$0.79$}&\multirow{2}{*}{$0.88$}&\multirow{2}{*}{$-0.81$}&\multirow{2}{*}{$0.74$} & \multirow{2}{*}{$0.43$}&0.75&0.080&0.215&$9.2$&$2.0$&$4.4$&$1.8$\\
&&&&&&&&&0.45&0.135&0.406&$5.5$&$1.0$&$3.7$&$1.0$\\
\hline
\multirow{2}{*}{CMS $B$ parking, $\mathcal{V} > 0.2 \;$} & \multirow{2}{*}{$8.4 \times 10^5$} & \multirow{2}{*}{$800$} & \multirow{2}{*}{$1.2 \times 10^6$}&\multirow{2}{*}{$0.83$}&\multirow{2}{*}{$0.82$}&\multirow{2}{*}{$-0.78$}&\multirow{2}{*}{$0.71$} & \multirow{2}{*}{$0.35$}&0.75&0.021&0.055& $> 10$&$6.4$&$4.9$&$3.9$\\
&&&&&&&&&0.45&0.036&0.110& $> 10$&$3.2$&$4.9$&$2.7$\\
\hline\hline
\end{tabular}
\caption{The $b\bar b$ cross section times the muon cuts efficiency, $\sigma\epsilon_{\mu\mu}$ (with the branching and fragmentation fractions factored out), integrated luminosity $\mathcal{L}$, number of expected signal events $N$ (after the full selection), expected values for the diagonal spin correlation matrix elements, the quantities $\Delta$ and $\mathcal{V}$, as well as their statistical uncertainties and significances. In the last two columns, we also show their total significance for a scenario with 20\% systematic uncertainty. We present the statistical uncertainties for two values of the polarization retention factors $r_i$ from Eq.~\eqref{eq:HadToParton Relation}: we fix $r_T=0.7$ and take an optimistic case of $r_L = 0.75$ in the first subrow and a pessimistic case of $r_L = 0.45$ in the second subrow. When a cut is applied to the expected values of $\Delta$ or $\mathcal{V}$, this is indicated in the first column.}
\label{tab:number}
\end{table*}

We generate $pp\to b\bar{b}$ events at NLO QCD using the \textsc{MadGraph5\_aMC@NLO}~\cite{Alwall:2014hca} Monte Carlo (MC) generator, using NNPDF3.1 NLO PDFs with $n_f=4$ and $\alpha_s(m_Z)=0.118$~\cite{NNPDF:2017mvq}. We then pass them through \textsc{Pythia}~8.3~\cite{Sjostrand:2014zea} for parton showering, hadronization and decays. We use the anti-$k_{\text T}$ algorithm~\cite{Cacciari:2008gp,Cacciari:2011ma} with $R=0.4$ for jet clustering and consider only the two leading-$p_{\text T}$ $b$-jets in each event.
Jets are required to be in the pseudorapidity range $2<\eta<5$ for LHCb, $|\eta|<2.4$ for Run~2 ATLAS and CMS, and $|\eta|<2.5$ for HL-LHC ATLAS and CMS. To estimate the event counts and spin correlations of the signal alone, we require the selection cuts on muons described below to be satisfied by muons produced directly in $b$-hadron decays and not in charmed-hadron decays or other sources, using truth-MC information. We then use the binned truth parton-level values of $M_{b\bar b}$ and $\cos\Theta$ in the selected events to analytically compute the $b\bar b$ spin correlation components for each bin. 
At NLO, in addition to $gg$ and $q \bar q$ initial states, $gq$ and $g \bar q$ also contribute.
However, their contribution was found to be negligible, so we omitted these events from our calculations.

In the analyses proposed for ATLAS, we rely on event selection via the double-muon trigger, which requires a pair of muons with $p_{\text T} > 15$~GeV and $|\eta|<2.4$ in Run~2 or $p_{\text T} > 10$~GeV and $|\eta|<2.5$ at the HL-LHC~\cite{ATLAS:2020gty,ATL-PHYS-PUB-2019-005}. Backgrounds from processes with prompt or mildly displaced muons (originating from charm or $\tau$ decays) are assumed to be eliminated by a $b$ tagging requirement, which needs to be satisfied by at least one of the jets, with an $\epsilon_b \approx 80\%$ efficiency for $b$ jets. Backgrounds from $c$-hadron decays in $b$ jets, where the muons are typically softer than those produced directly in $b$-hadron decays, are assumed to be eliminated by applying the requirement $p_{\text T}^\mu/p_{\text T}^{\rm jet} > 0.2$ to at least one of the muons. Finally, to reduce the background due to the semileptonic $B$-meson decays, we require one of the two jets to contain a fully reconstructed $\Lambda_c^+$ baryon (via one of its fully reconstructible decay modes, such as $\Lambda_c^+ \to pK^-\pi^+$)~\cite{SM}. We estimate the number of signal events that will be available for the analysis as
\begin{equation}
\begin{aligned}
    N =\;& 2\,\sigma\epsilon_{\mu\mu}\,\mathcal{L}\, f^2(b\to \Lambda_b)\, \text{BR}^2(\Lambda_b\to X_c \mu^- \bar{\nu}_\mu)\\
    &\times \text{BR}(\Lambda_c^+\to\text{reco.})\,\epsilon_{\text{reco.}}\,\epsilon_{b,2} \,,
\end{aligned}
\label{eq:Number_events}
\end{equation}
where $\mathcal{L}$ is the integrated luminosity and $\sigma\epsilon_{\mu\mu}$ is the $b\bar b$ production cross section times the muon cuts efficiency, with the branching and fragmentation fractions factored out. We take the fragmentation fraction to be $f(b\to\Lambda_b) \approx 7\%$ for $\Lambda_b$ with $p_{\text T} > 25$~GeV~\cite{Galanti:2015pqa}. For $p_{\text T} < 25$~GeV, we implement a $p_{\text T}$-dependent enhancement of the $\Lambda_b$ fragmentation fraction~\cite{LHCb:2019fns,LHCb:2023wbo} based on the results of Ref.~\cite{LHCb:2019fns}. We take the branching fractions $\mbox{BR}(\Lambda_b\to X_c\mu^-\bar{\nu}_\mu) \approx 11\%$ and $\text{BR}(\Lambda_c^+\to\text{reco.}) \approx 18\%$~\cite{ParticleDataGroup:2022pth}, assuming the list of $\Lambda_c^+$ decay channels included in Ref.~\cite{Kats:2023zxb}. The factor $\epsilon_{\text{reco.}} \approx 50\%$ is our rough estimate for the average $\Lambda_c^+$ decay reconstruction efficiency, and $\epsilon_{b,2} \equiv 2\epsilon_b - \epsilon_b^2$ is the efficiency for at least one of the two jets to pass the $b$-tagging condition.

The CMS $B$ parking dataset of Run~2~\cite{Bainbridge:2020pgi,CMS:2024zhe,CMS:2024syx,CMS:2024ita} is based on a trigger requiring a displaced muon with $|\eta|<1.5$ and $p_{\text T}$ above thresholds between $7$ and $12$~GeV. The information on the integrated luminosity for each $p_{\text T}$ threshold is available in Refs.~\cite{CMS:2024syx,CMS:2024zhe}. Our proposed analysis requires the presence of an additional muon (in another jet) with an opposite charge, $p_{\text T} > 5$~GeV and $|\eta|<2.4$. The advantage of this dataset is the higher statistics thanks to the low $p_{\text T}$ thresholds relative to the typical ones used in CMS and ATLAS. We calculate the number of expected signal events accounting for the fact that a muon is already required on one side:
\begin{equation}
\begin{aligned}
    N =&\, 2f^2(b\to\Lambda_b)\,\text{BR}(\Lambda_b \to X_c\mu^-\bar\nu_\mu)\,\epsilon_{\mu_2}\,\\
    & \times \text{BR}(\Lambda_c^+ \to {\rm reco.})\, \epsilon_{\rm reco}\,N_0 \;,
\end{aligned}
\end{equation}
where $N_0 \approx 10^{10}$ is the number of $b\bar b$ events in the CMS $B$ parking dataset~\cite{Bainbridge:2020pgi,CMS:2024zhe,CMS:2024syx,CMS:2024ita}, and $\epsilon_{\mu_2} \approx 36\%$ is the efficiency of selecting the muon on the nontriggering side of the event (found by our simulation). We also estimate the expected reach for the HL-LHC, assuming that such a trigger will be implemented. For concreteness, we  assume an integrated luminosity of $\mathcal{L}=800~\text{fb}^{-1}$ and rescale $N_0$ also by the ratio of our simulated cross section times efficiency at $\sqrt{s}=14$ vs.\ $13$~TeV.

We also consider LHCb, which has lower integrated luminosity than ATLAS or CMS, but also lower trigger thresholds and better reconstruction capabilities. Motivated by Refs.~\cite{LHCb:2017avl,LHCb:2022sck}, we consider a trigger requiring a muon with $p_{\text T} > 1.8$~GeV and $2 < \eta < 5$, a \mbox{two-,} three- or four-track secondary vertex with a significant displacement from any primary vertex (consistent with a $b$-hadron decay), with at least one charged particle with $p_{\text T} > 1.6$~GeV inconsistent with originating from a primary vertex. We simulate only the muon $p_{\text T}$ and $\eta$ requirements, assuming that the rest of the conditions will be satisfied for $b\bar b$ events with an efficiency close to $1$. Our proposed analysis requires the presence of a second muon (in another jet) with an opposite charge, $p_{\text T} > 0.5$~GeV and $2 < \eta < 5$. The number of expected signal events is computed with Eq.~\eqref{eq:Number_events}.

Our results are presented in Table~\ref{tab:number}. In some of the cases, as indicated in the table, we applied an additional cut on the expected parton-level value of $\Delta$ or $\mathcal{V}$ for the event to increase the significance of $\Delta$ or $\mathcal{V}$ in the sample. In these cases, for LHCb we also required $M_{b\bar b} > 20$~GeV at the parton level to focus on the relativistic entangled regime. 

\begin{figure}
\includegraphics[width=0.239\textwidth,trim=2mm 0 0 0,clip]{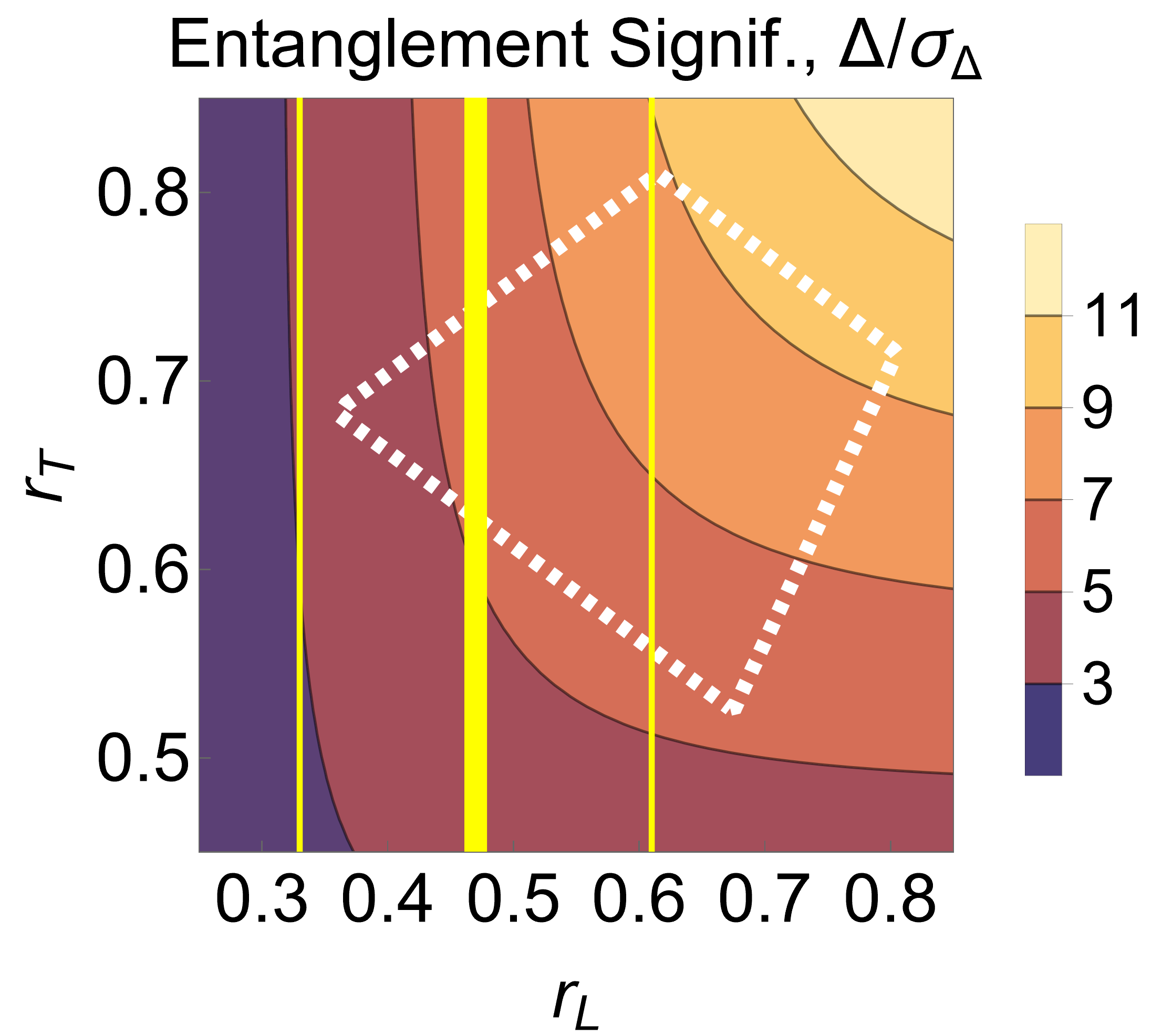}
\includegraphics[width=0.237\textwidth,trim=0 0 1.5mm 0,clip]{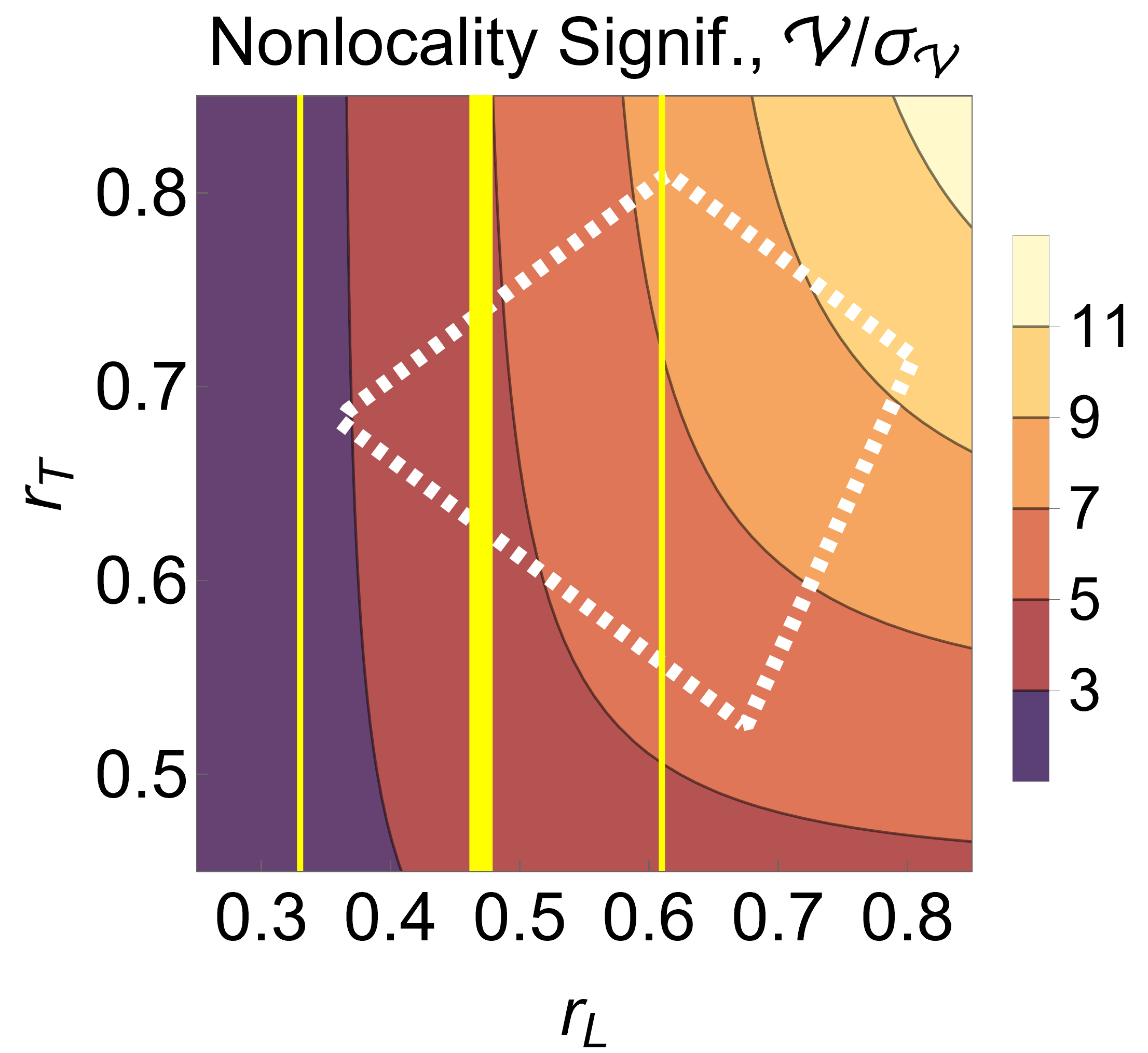}
\caption{The expected statistical significance of entanglement in Run~2 CMS $B$ parking dataset with the $\Delta > 0.2$ cut (left) and Bell nonlocality in the ATLAS HL-LHC dataset with the $\mathcal{V} > 0.3$ cut (right) as a function of the polarization retention factors $r_L$ and $r_T$. The white dotted polygons approximately indicate the region of plausible values for $r_L$ and $r_T$~\cite{Falk:1993rf,Galanti:2015pqa,SM}. The vertical yellow lines show the central value of $r_L$ (thick line) and its $\pm 1\sigma$ uncertainties (thin lines) from an approximate combination of the LEP measurements~\cite{ALEPH:1995aqx,OPAL:1998wmk,DELPHI:1999hkl}.}
\label{fig:significance}
\end{figure}

We find that the CMS $B$ parking dataset is the most promising for detecting entanglement in Run~2 data, with a statistical significance of above $5\sigma$, the exact number depending on the values of the polarization retention factors $r_L$ and $r_T$. This leaves room for obtaining high significance even after accounting for the systematic uncertainties. At the HL-LHC all the experiments are promising for detecting entanglement with high significance, and ATLAS and CMS show potential for detecting Bell nonlocality as well. In Fig.~\ref{fig:significance} we show how the statistical significance depends on $r_L$ and $r_T$ for the measurement of entanglement with the CMS $B$ parking dataset of Run~2 and for the measurement of Bell nonlocality with the ATLAS HL-LHC dataset.

\textit{Conclusions and discussion.---}We propose methods to measure entanglement and Bell nonlocality at the LHC with pairs of $b \bar b$ quarks.
This system is especially attractive given the large cross section for ultrarelativistic bottom quarks at the LHC.
We find that the observation of entanglement is possible with high significance with the currently available CMS $B$ parking data. Bell nonlocality is still beyond our current reach, but will be accessible using the full HL-LHC data. 
We find that, in addition to ATLAS and CMS, entanglement observation will also be possible using LHCb data, further motivating the study of quantum correlations at this detector.

In future work, it will be useful to extend the predictions for $b\bar b$ spin correlations beyond LO QCD. It will also be interesting to explore the possibility of measuring entanglement and Bell nonlocality using bottom quarks in future colliders, such as the FCC-ee and the FCC-hh~\cite{FCC:2018evy,FCC:2018vvp,Benedikt:2020ejr}.
Furthermore, additional QM concepts, such as discord and steering~\cite{Afik:2022dgh}, can also be addressed in $b \bar b$ pairs. 
In general, our work paves the way to study quantum correlations in hadronizing systems. Among others, this could have impact on the characterization of the quark-gluon plasma, whose vortical structure can lead to nontrivial spin properties~\cite{STAR2017,ALICE:2021pzu,HADES:2022enx,STAR:2022fan,Giacalone2025}.
The direct access to the ultrarelativistic regime provided by bottom quarks can also be of interest for the study of the relativistic behavior of the spin operator, a fundamental question in QM that is still open~\cite{Czachor1997,Gingrich2002,Peres2004,Friis2013,Rembielinski2019,Giacomini2019,Taillebois2021,Kurashvili2022}. Finally, it could be interesting to consider even lighter quark-antiquark systems, such as $c\bar c$ or $s\bar s$, which however seem more challenging~\cite{Kats:2023zxb}.

\medskip
\acknowledgments{YA is supported by the National Science Foundation under Grant No.\ PHY-2310094. YK and DU are supported in part by the Israel Science Foundation (Grant No.~1666/22) and the United States~- Israel Binational Science Foundation (Grant No.~2018257). JRMdN is supported by European Union's Horizon 2020 research and innovation programme (Marie Sk\l{}odowska-Curie Grant Agreement No.\ 847635), and by Spain's Agencia Estatal de Investigaci\'on (Grant No.\ PID2022-139288NB-I00).  AS is supported by the Israel Science Foundation (grant Nos.~203/23 and 3464/21), the United States-Israel Binational Science Fund (grant No.~2020044), and the Horizon 2020 Marie Sklodowska-Curie RISE project JENNIFER2 (grant No.\ 822070).}

\section*{Supplemental Material}

\subsection*{Polarization retention factors}

As analyzed in \cite{Falk:1993rf} and developed further in \cite{Galanti:2015pqa}, in the heavy-quark limit, the polarization retention factors in $b\to\Lambda_b$ fragmentation can be expressed in terms of two nonperturbative QCD parameters, $A$ and $w_1$, as
\begin{eqnarray}
r_L \approx \frac{1+A\left(0.23 + 0.38 w_1\right)}{1+A} \,, \\
r_T \approx \frac{1+A\left(0.62 - 0.19 w_1\right)}{1+A} \,.
\end{eqnarray}
These quantities, $r_L$ and $r_T$, are the fractions of the initial $b$-quark polarization (in the longitudinal and transverse directions, respectively, relative to the fragmentation axis, i.e., the $b$ quark momentum direction) that are retained in the final $\Lambda_b$ polarization. The above expressions describe the dominant polarization loss effect, due to the contribution to the $\Lambda_b$ sample from $\Sigma_b^{(\ast)} \to \Lambda_b \pi$ decays. The parameter
\begin{equation}
A = \frac{\mbox{prob}\,(\Sigma_b^{(\ast)})}{\mbox{prob}\,(\Lambda_b)} = 9\,\frac{\mbox{prob}\left(T\right)}{\mbox{prob}\left(S\right)}
\end{equation}
is the ratio of the $\Sigma_b^{(\ast)}$-decay and direct $\Lambda_b$ production rates.
It is related to the probability for the two light quarks in the baryon to form any of the nine spin-triplet, isospin-triplet diquark states $T$ and the probability to form the spin-singlet, isospin-singlet diquark state $S$. The statistical hadronization model (for a brief overview, see \cite{Andronic:2009sv}) predicts $A \approx 2.6$~\cite{Galanti:2015pqa}, but it is unclear how accurate this number is. The parameter
\begin{equation}
w_1 = \frac{\mbox{prob}\,(T_{\pm 1})}{\mbox{prob}\,(T)}    
\end{equation}
accounts for the possibility that the fragmentation axis breaks the
rotational symmetry in the spin-triplet diquark production. It describes the probability for the diquark to be produced with spin component $+1$ or $-1$ (but not $0$) along the fragmentation axis. The isotropic case is obtained for $w_1 = 2/3$. The value of $w_1$ can be determined from the angular distributions of the pions in the $\Sigma_b^{(\ast)} \to \Lambda_b \pi$ decays~\cite{Falk:1993rf,Galanti:2015pqa}. Measurements of both $A$ and $w_1$ can certainly be done at LHCb~\cite{LHCb:2018haf} and perhaps even at ATLAS and CMS. The white dotted polygons in Fig.~2 of the main text correspond to the range
\begin{equation}
    1 \leq A \leq 5 \,,\qquad
    0 \leq w_1 \leq 1 \,,
\end{equation}
where the chosen range for $A$ reflects a large systematic uncertainty.

\subsection*{Background due to semileptonic $B$-meson decays}

An important background to the semileptonic $\Lambda_b$ decays is due to semileptonic decays of $B$ mesons. Reference~\cite{Galanti:2015pqa} proposed three possible approaches to dealing with this background. The first approach (``Inclusive Selection'') does not attempt to reduce it. This results in low sample purity, but keeps the signal efficiency high. The $B$-meson background can be reduced (with a corresponding cost in signal statistics) by requiring the jet to contain a reconstructed $\Lambda_c^+$ baryon (via one of its fully reconstructible decay modes, such as $\Lambda_c^+ \to pK^-\pi^+$) or a $\Lambda$ baryon (reconstructed via its $\Lambda \to p\pi^-$ decay). These are referred to as ``Exclusive Selection'' and ``Semi-Inclusive Selection'', respectively~\cite{Galanti:2015pqa}. Each of these requirements can be applied to either one or both sides of the event, leading to six possible analysis channels, all of which were analyzed in \cite{Kats:2023zxb}. The statistical uncertainty was found to be the lowest for the inclusive/exclusive channel (although the other channels turned out rather comparable). The sample purity obtained in this channel is $\sim 4.9\%$. A higher sample purity of $\sim 44\%$ is possible in the exclusive/exclusive channel at the price of lowering the signal efficiency by a factor of $\sim 22$ and increasing the statistical uncertainty for the spin correlation components by a factor of $\sim 1.6$. We base our estimates in this Letter on the inclusive/exclusive channel, while noting that sensitivity can be improved by combining all six channels or by utilizing electrons in addition to muons, which we do not pursue here.

\subsection*{Statistical uncertainty estimation}

The expected statistical uncertainty in a measurement of the spin correlation matrix components $C_{ij}$, using a fit of the data to Eq.~(3) from the main text and the relation in Eq.~(4) therein, is approximately~\cite{Kats:2023zxb}
\begin{equation}
    \sigma_{C_{ij}}^{\rm stat} \simeq \frac{3}{r_i r_j\alpha^2\sqrt{fN}} \;,
\end{equation}
where $N$ is the expected number of signal events after the full event selection, $f$ is the sample purity $N/(N+N_B)$, with $N_B$ being the number of background events, and we have approximated the angular distribution of the background to be similar to that for $C_{ij} = 0$.

\bibliography{main.bib}

\end{document}